\documentclass[preprint2]{aastex}
\usepackage{lscape}
\usepackage{epsf}
\usepackage{graphics}

\newcommand{\kms}{km s$^{-1}$}
\newcommand{\cts}{cts s$^{-1}$}
\newcommand{\ax}{$\alpha_{\rm X}$}
\newcommand{\aox}{$\alpha_{\rm ox}$}
\newcommand{\cm}{cm$^{-2}$}
\newcommand{\rb}[1]{\raisebox{1.5ex}[-1.5ex]{#1}}

\newcommand{\pl}{$\pm$}
\newcommand{\nh}{$N_{\rm H}$}
\newcommand{\chandra}{{\it Chandra}}
\newcommand{\newton}{{\it Newton}}
\shorttitle{XMM Observations of BAL QSOs}
\shortauthors{Grupe et al.}

\begin{document}

\def\clipfig#1{\def\lbracket{[}\def\testit{#1}%
    \ifx\testit\lbracket\let\next=\optclipfig\else\let\next=\stdclipfig\fi%
    \next{#1}}
%
\newcommand {\hclipfig} [7] {\clipfig[#7]{#1}{#2}{#3}{#4}{#5}{#6}}
%
\def\usemodepsfig {\global\def\cfmode{x}\typeout{*** set clipfig to PSFIG mode ***}}
\def\usemodeepsf  {\global\def\cfmode{}\typeout{*** set clipfig to EPSF mode ***}}
\def\useunitmm    {\global\def\cfunit{x}\typeout{*** set clipfig to use mm as unit ***}}
\def\useunitcm    {\global\def\cfunit{}\typeout{*** set clipfig to use cm as unit ***}}
\def\clipfigsettings {\ifx\cfmode\empty\def\ccfmode{EPSF }\else\def\ccfmode{PSFIG }\fi%
    \ifx\cfunit\empty\def\ccfunit{cm }\else\def\ccfunit{mm }\fi%
    \typeout{*** current clipfig settings: \ccfmode mode, using \ccfunit as unit ***}}
%
%
%
%
\def\stdclipfig#1#2#3#4#5#6{\ifx\cfmode\empty%
    \let\next=\eclipfig\else\let\next=\pclipfig\fi%
    \next{#1}{#2}{#3}{#4}{#5}{#6}}
\def\optclipfig#1#2]#3#4#5#6#7#8{\ifx\cfmode\empty%
    \let\next=\ehclipfig\else\let\next=\phclipfig\fi%
    \next{#3}{#4}{#5}{#6}{#7}{#8}{#2}}
%
%
%
\newcommand {\pclipfig}[6] {\ifx\cfunit\empty%
        \psfig{figure=#1.ps,width=#2cm,bbllx=#3cm,bblly=#4cm,bburx=#5cm,%
           bbury=#6cm,clip=}\else%
        \psfig{figure=#1.ps,width=#2mm,bbllx=#3mm,bblly=#4mm,bburx=#5mm,%
           bbury=#6mm,clip=}\fi}
\newcommand {\phclipfig}[7] {\ifx\cfunit\empty%
        \hspace{#7cm}\psfig{figure=#1.ps,width=#2cm,bbllx=#3cm,bblly=#4cm,%
           bburx=#5cm,bbury=#6cm,clip=}\else%
        \hspace{#7mm}\psfig{figure=#1.ps,width=#2mm,bbllx=#3mm,bblly=#4mm,%
           bburx=#5mm,bbury=#6mm,clip=}\fi}
%
%
%
\newcommand {\eclipfig}[6]{%
  \ifx\cfunit\empty\epsfxsize=#2cm\else\epsfxsize=#2mm\fi%
  \epsfclipon\epsfverbosetrue%
  \cfcmtopspts{#3}\cfllxi=\cftempi\cfllxf=\cftempf%
  \cfcmtopspts{#4}\cfllyi=\cftempi\cfllyf=\cftempf%
  \cfcmtopspts{#5}\cfurxi=\cftempi\cfurxf=\cftempf%
  \cfcmtopspts{#6}\cfuryi=\cftempi\cfuryf=\cftempf%
  \def\cfstra{\number\cfllxi.\number\cfllxf}%
  \def\cfstrb{\number\cfllyi.\number\cfllyf}%
  \def\cfstrc{\number\cfurxi.\number\cfurxf}%
  \def\cfstrd{\number\cfuryi.\number\cfuryf}%
  \hbox{\epsfbox[{\cfstra} {\cfstrb} {\cfstrc} {\cfstrd}]{#1.ps}}}
\newcommand {\ehclipfig}[7]{%
  \ifx\cfunit\empty\epsfxsize=#2cm\else\epsfxsize=#2mm\fi%
  \epsfclipon\epsfverbosetrue%
  \cfcmtopspts{#3}\cfllxi=\cftempi\cfllxf=\cftempf%
  \cfcmtopspts{#4}\cfllyi=\cftempi\cfllyf=\cftempf%
  \cfcmtopspts{#5}\cfurxi=\cftempi\cfurxf=\cftempf%
  \cfcmtopspts{#6}\cfuryi=\cftempi\cfuryf=\cftempf%
  \def\cfstra{\number\cfllxi.\number\cfllxf}%
  \def\cfstrb{\number\cfllyi.\number\cfllyf}%
  \def\cfstrc{\number\cfurxi.\number\cfurxf}%
  \def\cfstrd{\number\cfuryi.\number\cfuryf}%
  \ifx\cfunit\empty\hspace{#7cm}\else\hspace{#7mm}\fi%
  \hbox{\epsfbox[{\cfstra} {\cfstrb} {\cfstrc} {\cfstrd}]{#1.ps}}%
  \vspace{-1mm}}
%
%
%
\newdimen\cfllxi \newdimen\cfllyi  \newdimen\cfurxi  \newdimen\cfuryi
\newdimen\cfllxf \newdimen\cfllyf  \newdimen\cfurxf  \newdimen\cfuryf
\newdimen\cftemp \newdimen\cftempi \newdimen\cftempf
\newdimen\cfpspoint \cfpspoint=1bp
%
%
%
\newcommand{\cfcmtopspts}[1]{\ifx\cfunit\empty%
  \cftemp=#1cm\else\cftemp=#1mm\fi%
  \multiply\cftemp10\divide\cftemp\cfpspoint%
  \cftempf=\cftemp\divide\cftemp10\cftempi=\cftemp\multiply\cftemp10%
  \advance\cftempf-\cftemp}
%
%
\def\cfmode{}\def\cfunit{}\clipfigsettings
%

\useunitmm

\def \charthoffset {\hspace{0.2cm}} \def \charthsep {\hspace{0.3cm}}
\def \chartvsepcap {\vspace{0.3cm}}
\def \chartvsep {\vspace{0.1cm}}
\newcommand{\putchartb}[1]{\clipfig{#1}{83}{0}{0}{275}{170}}
\newcommand{\putchartc}[1]{\clipfig{#1}{83}{0}{0}{275}{195}}
\newcommand{\chartlineb}[2]{\parbox[t]{18cm}{\noindent\charthoffset\putchartb{#1}\charthsep\putchartc{#2}\chartvsep}}
\newcommand{\chartlinec}[2]{\parbox[t]{18cm}{\noindent\charthoffset\putchartc{#1}\charthsep\putchartc{#2}\chartvsep}}

\title{XMM-Newton Observations of Two BAL QSOs: Q1246-057 and SBS1542+541
\thanks{Based on
observations obtained with XMM-Newton, an ESA science mission with instruments
and contribution directly funded by ESA member states and NASA}
}


\author{D. Grupe, S. Mathur}
\affil{Astronomy Department, Ohio State University,
    140 W. 18th Ave., Columbus, OH-43210, U.S.A.}
\email{dgrupe, smita @astronomy.ohio-state.edu}

\author{and M. Elvis}
\affil{Harvard-Smithsonian Center for Astrophysics, 60 Garden Street,
    Cambridge, MA 02138, U.S.A.}




\begin{abstract}
We report on the results of XMM-Newton observations of two Broad
Absorption Line Quasars (BAL QSOs), Q1246-057 and SBS
1542+541. Unprecedented sensitivity of XMM allows spectral analysis of
these X-ray weak sources. The X-ray spectral data of these sources can
be fitted by a power-law with \ax=1.0-1.2 and either a partially
covering absorber or an ionized absorber model.  Rest frame UV
spectroscopy together with polarimetry favors the model with a
partially covering absorber with column density a few times
$10^{22}$\cm and a covering fraction of about 0.80. After correcting for
absorption, the X-ray loudness of these BAL QSOs appears to be similar
to other, unabsorbed quasars. The mystery of X-ray weakness of BAL QSOs
appears to be all but solved, with strong absorption being the primary
reason. With the available X-ray data, however, the issue of whether
BAL QSOs represent highly accreting/ younger population of quasars
remains unsettled.

\end{abstract}


\section{Introduction}

About 10-15\% of optically-selected Quasars have rest frame 
ultraviolet spectra showing deep broad absorption line (BAL) troughs
displaced blueward from the corresponding emission lines. 
Recently, \citet{hew03} found that the intrinsic fraction of BAL quasars 
(BAL QSOs) is as large as
22\% in the 1.5$\leq z \leq$3.0 redshift range.
These broad
absorption lines are seen either in high ionization transitions such as C~IV,
Si~IV, N~V, and O~VI or in low-ionization stages such as Al~III or
Mg~II (\citet{wey91}), dividing them into high-ionization and
low-ionization broad absorption line quasars (HiBAL and LoBAL).  If the broad
absorption lines are associated with mass outflows from the nucleus,
then outflow velocities of several thousand km s$^{-1}$ are inferred
(\citet{ara97}, but see \citet{branch02} for an alternative
explanation).  If quasar outflows are a result of radiation pressure
from the nuclear continuum, the large momentum associated with the
BAL QSO outflow causes a serous problem to the quasar energy budget
(see \citet{mat01} and references there in). As an alternative, outflows in AGN
can also be the result of disk-driven hydromagnetic winds (e.g. \citet{emm92,
koe94, bot97, bot00}).
 BAL QSOs in general show high
degrees of polarization, of up to more than 10\%, in their absorption
troughs (e.g. \citet{sch99, ogl99}). This argues that part of the
nuclear emission is scattered back into our line of sight off a highly
asymmetric structure. BAL QSOs may then be normal quasars viewed down
the 10--20\% of the lines of sight covered by ubiquitous accelerating
wind (e.g. Elvis 2000)

While these properties make BAL QSOs important in their own right, some
recent observations suggest that they also occupy a special space in
quasar life line (e.g. \cite{mat00}, \citet{bec00}).  BAL QSOs,
together with narrow line Seyfert 1 galaxies (NLS1s), occupy one
extreme end of the Eigenvector 1 for quasars and Seyferts
(\cite{bor02}). As such, they appear to be objects accreting at close
to the Eddinton limit. NLS1s have been suggested to be AGN in a young
state of their evolution (\citet{gru96, gru99, mat00}).  An analogy
between BAL QSOs, especially the low-ionization BAL QSOs, and NLS1s
(\citet{bra00, law97, lei97, bra97}) further suggests that BAL QSOs possibly
are young quasars (\citet{mat00, bec00}).
This possibility adds further
significance to the study of BAL QSOs. 

X-ray studies of BAL QSOs are important for many reasons. X-ray
observations offer direct measurements of the total absorbing column
density.  In the last decade the rest-frame UV spectra of BAL QSOs
have been extensively investigated.  However, using the UV range only,
properties such as absorption column density, covering fraction, or
temperature are poorly constrained. In fact, absorption columns
derived from the UV absorption lines led to values of
$N_H~=~10^{20}~-~10^{21}$\cm~(\citet{kor92}). This would suggest only
minor intrinsic X-ray absorption in these sources. However, BAL QSOs
appear to be highly absorbed in soft X-rays, implying absorbing column
densities at least an order of magnitude higher compared to those from
the UV studies (\cite{gre01} and references there in). Scattering into
our line of sight, evidenced by polarization (\citet{ogl99}) turns
these measurements into lower limits, and allows consistency with the
10--100 times larger X-ray column densities. Non-black saturation of
absorption lines (e.g. \cite{ara97}) is a probable cause of the
discrepancy between the X-ray and UV studies. Another motivation
behind X-ray observations of BAL QSOs is to determine the intrinsic
spectral shape in the X-ray band and to compare it with the NLS1
spectra. Are the X-ray spectra of BAL QSOs similar to those of non-BAL
QSOs?

New X-ray missions, \chandra~ and XMM-\newton, offer complementary
advantages to BAL QSO studies. \chandra, with its excellent point
spread function and low background is ideal for detecting faint
sources like BAL QSOs.  XMM-\newton~ (\citet{jan01}), on the other hand, 
with its large
collecting area, is ideal for obtaining spectra. In a recent survey of
BAL QSOs with \chandra, \citet{gre01} detected 8 out of the 10
observed BAL QSOs (see also \cite{gal02a}). In this paper we report on
XMM-\newton~ observations of two BAL QSOs, Q1246-057 and SBS
1542+541. Both the objects were previously detected in X-rays, and
given their flux, we expected to obtain a good S/N spectrum with
XMM-\newton. Unfortunately, QSO Q1246-057 was observed to be a factor
of two fainter compared to epoch of \chandra~ observation and about
half of SBS 1542+541 data were unusable because of high background
radiation. As a result, the spectral quality of both objects is not as
good as expected, but given the rarity of BAL QSO spectra, any spectral
information is valuable. In the following we briefly discuss the
properties of our two targets and then proceed to the discussion of
observations ($\S 2$), results ($\S 3$), discussion ($\S 4$) and we
conclude in $\S 5$.

Q1246-057 ($\alpha_{2000}$=12h 49m 13.9s,
$\delta_{2000}=-05^{\circ}~59^{'}~19.3{''}$, z=2.236,
galactic \nh=2.15 10$^{20}$ \cm~ 
(\citet{dic90})) is a
high-ionization BAL QSO discovered by \citet{osm77}. They
found strong, broad (5000 \kms) absorption lines which were
blue-shifted by 15,000 \kms~ with respect to the quasars rest
frame. The source was re-observed with higher spectral resolution by
\citet{bok78} showing multicomponent structure of the absorption
troughs. In spectropolarimetry measurements \citet{sch99} and
\citet{ogl99} found strong polarization in the C~IV~$\lambda$1550
trough, with increasing degree of polarization toward the blue, of
the order of about 1.1\% to 2\% (\citet{hut98, sch99, ogl99}). X-ray
observations with ROSAT and \chandra~
(\citet{pag00, pag01, gre01})
detected the source, but did not yield enough photons to perform
spectral analysis.

SBS~1542+541 ($\alpha_{2000}$=15h 43m 59.4s,
$\delta_{2000}=+53^{\circ}59^{'}03.7^{''}$, z=2.361,
galactic \nh=1.27 10$^{20}$ \cm~
 (\citet{dic90})) has shown a
variety of highly ionized absorption lines in HST spectra blue-shifted
by $\approx$ 11000-12000 \kms~ and widths of 2000-3000 \kms
(\citet{tel98}). It was detected by ROSAT (\citet{yua98} with
7.02$\times~10^{-3}$ \cts~(\citet{tel98}) in the Position Sensitive
Proportional Counter (PSPC, \citet{pfe86}). In a later observation
with \chandra's ACIS-S detector the quasar was clearly detected with a count
rate of 0.019 \cts~(\citet{gre01}).

Throughout the paper we use energy spectral slopes $\alpha$
defined as $F_{\nu} \propto \nu^{-\alpha}$. Luminosities are calculated
assuming a Hubble constant of $H_0$ =75 \kms Mpc$^{-1}$ and a
deceleration parameter of $q_0$ = 0.5 if not noted otherwise.

\section{\label{observe} Observations}

\subsection{\label{xmm_observe} XMM-Newton observations}

Q1246-057 was observed by XMM 
for 41.5 ks in orbit 291.  SBS 1542+541
was observed twice in orbits 294 and 396 for 11 and 36 ks,
respectively.  The observations of both sources were performed with
XMM-Newton's EPIC PN and MOS cameras (\citet{str01} and \citet{tur01},
respectively)
in Full-Frame mode using thin
filters (Table \ref{xmmobs}).  For Q1246--057 the whole lengths of the
observations could be used. Due to high background radiation in some
parts of the observations of SBS 1542+541 only part of these
observations were used.  Only times with high energy ($>$ 10 keV)
background count rate CR $<$ 10 \cts~ were accepted and screened into
good time intervals (GTIs). These GTIs were used to create new event
file lists that contain only acceptable data. In the end, the two GTI
event files were merged into one which was used to derive
spectra. This results in total observing time of 25176 s for the PN
and 37303 s for the MOS detectors.  

Source photons were selected in a
circle with a radius of 22.5$^{''}$ and 25.0$^{''}$ in the PN and MOS,
respectively. Background photons were selected for a circular region
of 50$^{''}$ radius close to the source on the same CCD. Only good
quality events, with single and double patterns for the EPIC PN (PATTERN$\leq$4)
and single, double, and triple events for the EPIC MOS (PATTERN$\leq$12), were
selected for spectral analysis. The count-rates of both sources are
too low to be affected by pile-ups (See Sect. \ref{q1246_res} and
\ref{sbs1542_res}).

The data were reduced by using the XMM-\newton~ Science Analysis
Software (XMMSAS) version 5.3.3 and the X-ray spectra were analyzed with
XSPEC 11.2.0. The spectra were grouped by tool GRPPHA 3.0.0 in bins of at
least 20 counts per bin.  For the count-rate conversions between
different X-ray missions, PIMMS 3.2 has been used.

\subsection{\label{rosat_observe} ROSAT observations}

Q1246--057 and SBS 1542+541 were both targets for pointed ROSAT
\citet{tru83} observations using the PSPC.
Q1246--057 was observed serendipitously in an
observation of NGC4697 on 1993-06-22 to 1993-07-07 (rp600262a02) for a
total of 45 ks (\citet{gre96, pag00, pag01}).  SBS 1542+541 was
observed on 1993-08-14 (rp701436) for a total of 5.6 ks
(\citet{tel98}). In both cases we re-analyzed the ROSAT data using the
Extended X-ray Scientific Analysis System (EXSAS, \citet{zim98})
version 01APR. These data were used together with the XMM data for
spectral analysis (see Sect. \ref{results}).

\subsection{Chandra Observations}

Both sources have been observed by the \chandra~X-ray observatory with 
the Advanced CCD Imaging Spectrometer (ACIS-S)
(\citet{gre01}). Q 1246--0542 was observed on 2000-02-08 for 5.4 ks.  An ACIS-S
count rate of 0.0081 \cts was measured. For SBS 1542+541 which was observed on
2000-03-22 a count rate of 0.0197 \cts~ was measured (\citet{gre01}).

\section{\label{results} Results}

\subsection{\label{q1246_res} Q1246--057}

\subsubsection{Long-term light curve}

The mean count rates measured for Q1246--057 by the EPIC PN, MOS-1,
and MOS-2 are (1.21\pl0.15) $\times~ 10^{-2}$ \cts, (3.86\pl0.41)
$\times~10^{-3}$ \cts, and (4.18\pl0.70) $\times~ 10^{-3}$ \cts,
respectively, resulting in 460, 160, and 170 photons. Due to the low
count rate of the source no reliable light curve could be derived.

 We measured the ROSAT PSPC count-rate to be
(1.30\pl0.29)$\times~10^{-3}$ \cts, consistent with that reported by
\citet{pag00}.  Using the EPIC PN count-rate and the best fit model
  ( table \ref{q1246_spec}) we calculated the ROSAT PSPC count rate
  using PIMMS and found it to be the same as observed. This suggests
  that the source did not vary significantly between the ROSAT and
  XMM-\newton~ observations. However a comparison to the Chandra ACIS-S
  count rate of CR=(8.1\pl1.2) 10$^{-3}$\cts~ 
  (\citet{gre01}) suggests that the source was about a
  factor of two brighter during the \chandra~ observation compared to
  the ROSAT and XMM observations. Converting the EPIC PN count rate into an 
  ACIS-S count rate gives CR= (4.9\pl0.5) 10$^{-3}$\cts.
   From the quality of the data it is
  not clear if this is due to intrinsic variability in luminosity of
  the source or caused by a change in the absorber property, either
  column density, covering fraction or the ionization parameter. The
  \chandra observation was performed on 2000-02-08,
so the Chandra and XMM observations were 520 days apart which
converts to 161 days in the rest-frame of the quasar. From the ROSAT observation
the Chandra and XMM observations were 2596 and 2934 days apart, respectively,
which converts to 802 and 907 days in Q 1246--057's rest-frame, respectively.

\subsubsection{Spectral analysis}

Figure 1 shows the observed XMM-\newton~ spectrum of the source in the
0.2--7 keV range ($\sim 0.6-21$ keV rest frame). What is immediately
apparent is lack of low-energy cut-off. This is completely contrary to
the expectation, based on earlier work, that BAL QSOs are highly
absorbed in X-rays (\cite{mat01, gre01}). A simple power-law model
with only Galactic absorption fits the data with $\chi^2=6.9$ for 12
degrees of freedom (Tab. \ref{q1246_spec}). If the absorbing column density
is allowed to vary, the fit yields \ax$=1.28\pm 0.94$, and $N_{\rm H}$
is unconstrained (PN data only).  If the source is not intrinsically
X-ray weak, and is absorbed in X-rays like other BAL QSOs, then the
observed flux in the soft-energy range can be explained if (1) the
absorber only partially covers the source, or (2) the absorber is
ionized. We examine these two alternatives below.

We then attempted to fit the data with an absorbed power-law model,
partially covering the continuum source. Such a model was motivated by
our earlier work on BAL QSOs (\citet{mat01}).  The S/N in the
spectrum is not good enough to perform complex spectral fits, and
leaving the spectral slope parameter free results in fits where no
parameter can be constrained (see Table \ref{q1246_spec}). Restricting
the EPIC PN data in the observed 1.5-7.5 keV energy range yields
\ax=1.28 when \nh~ is free (Table \ref{q1246_spec}). This energy range
converts to 4.85-24.3 keV in the rest-frame of Q1246--057 which is
insensitive to absorption with column densities smaller than $\approx$
\nh=10$^{23}$\cm~ and
therefore the X-ray spectral slope \ax~ in this range is likely the
intrinsic un-affected power-law slope of the source. We therefore
fixed the power-law slope to \ax=0.80, 1.0, 1.28 (to match the best
fit), and 1.5 (See Table \ref{q1246_spec}).

Fitting all three XMM EPIC detectors simultaneously results
in a partially covering absorber with
\nh=(4.49\pl2.36)$\times~10^{22}$ \cm~ with a covering fraction of
58\pl9\% (Table \ref{q1246_spec}) for \ax=1.0 (fixed) and
\nh=(7.4\pl2.1)$\times~10^{22}$ \cm~ with a covering fraction of 79\%
(Table \ref{q1246_spec}) for \ax=1.5 (fixed). Addition of the ROSAT
data do not contribute significantly to the fit. However Table
\ref{q1246_spec} shows that the ROSAT PSPC spectrum can be fitted with
the same parameters as the EPIC PN and MOS data, providing a
consistency check. 

The rest frame UV spectrum of Q1246--057 shows absorption lines of
 Si\,IV$\lambda$1397 and C\,IV$\lambda$1549 (\citet{bok78}, their
 figure 2) in which the flux in troughs do not reach zero
 intensity. If this is due to non-black saturation resulting from
 partial covering of the continuum source, then the estimated covering
 fraction is 0.8. This is consistent with covering fraction determined
 with the X-ray spectral fitting, for \ax=1.5 (Table
 \ref{q1246_spec}).

As discussed above, an ionized absorber can also be transparent in the
soft X-ray region. To test this possibility we fitted the spectrum
with an ionized absorber model in XSPEC, and the results are given in
\ref{q1246_spec}. Again, because the data quality is not good enough to fit
all the parameters simultaneously, we fixed the power-law slope to
\ax=1.0 and then to \ax=1.5 to fit the absorber parameters. As shown
in Figure \ref{q1246_plot_cont}, the column density and ionization
parameter of the absorber are not well constrained, but the best fit
values are $N_{\rm H}=6.5\pm1.9 \times 10^{22}$ cm~ and the ionization
state (\citet{don92}) $\xi=609\pm175$ for \ax=1.0. For \ax=1.5, the
best fit values are $N_{\rm H}=6.9\times 10^{22}$\cm and
$\xi$=1000. In terms of the more familiar representation of the
ionization parameter U (\citet{fer98}), these best fit values
correspond to U=30 and 50, respectively, using the conversion given in
Fig. 1 in \citet{geo98}. This compares well with the ionization
parameter U$\approx$20 suggested by \citet{sab01} for the BAL QSO PG
1254+047 derived from \chandra~ observations. However, compared to
ionization parameters U derived from UV observations of BAL QSOs, which are in
the order of 0.01-5
(e.g. \citet{ham98, tur96}), this value is rather large.

 We determined the X-ray loudness \aox\footnote{\aox~
is the slope of a hypothetical power-law from 2500 \AA~ to 2 keV;
\aox=0.384$\log~(L_{2500}/L_{2keV})$}  using
 the unabsorbed rest-frame fluxes $F_{\rm 2500\AA}$ and $F_{\rm
 2keV}$. The $F_{\rm 2500\AA}$ flux was determined by extrapolating
 the continuum in the spectrum given in Fig.  4 in \citet{sch99} to
 2500\AA~ and adding 20\% additional flux density due to flux losses,
 e.g. slit losses (G. Schmidt priv. comm). The unabsorbed rest-frame
 2keV X-ray flux density was determined from the the EPIC PN
 spectrum. Using these data we determined \aox=1.50 (Table
 \ref{xraylum}). 

\subsection{\label{sbs1542_res} SBS 1542+541}

\subsubsection{Long-term light curve}

In the EPIC PN and MOS-1 and MOS-2 observations SBS 1542+541 detected
count-rates of (4.51\pl0.21) $\times10^{-2}$ \cts,
(1.38\pl0.08)$\times 10^{-2}$ \cts, and (1.30\pl0.09)$\times 10^{-2}$
\cts, resulting in 1135, 515, and 485 total photons respectively. Due
to the presence of high background during part of the observation, the
observed data quality is not as good as expected and it is impossible
to perform reliable temporal analysis. We did not detect any
significant change in count rate between the two observations four
days apart.

We measured the ROSAT PSPC
count-rate CR = (5.19\pl1.15) $\times 10^{-3}$ \cts. Using PIMMS
and a power-law model with partial covering with all parameters free
as given in Table \ref{sbs1542_spec} results in a similar count-rate
(5.06$\times10^{-3}$\cts). Using the same data to derive a count-rate
for the \chandra~ ACIS-S results in 0.016\pl0.001 \cts, which agrees with the
 measured value (0.019$\pm0.002$ \cts, \citet{gre01}). The \chandra~ and
 XMM-\newton~ observations were separated by
 2412 and 3097 days from the ROSAT observation, which converts to 718 and 921
 days in the rest-frame. The Chandra and XMM observations were separated by
 686 days in the
 observed frame or 204 days in the rest-frame.

\subsubsection{Spectral Analysis}

Spectral analysis of SBS 1542+541 was performed the same way as
discussed above for Q 1246-057, and the results are summarized in
table \ref{sbs1542_spec}. When leaving all parameters free (except the
Galactic absorption) the X-ray slope \ax~ is of the order of 1.0 in
all the fits and the covering fraction is of the order of 0.70. The
absorption column is constrained  to lie between about 2.7 to 6.4 $\times
10^{22}$ cm$^{-2}$ (see also
Fig. \ref{sbs1542_plot}). A combined fit to the spectra from all the EPIC
detectors results in power-law slope \ax=1.04 with an intrinsic
absorption column \nh$\approx~4\times10^{22}$\cm~ and a covering
fraction of $\approx$0.70. Fixing the power-law slope to \ax=1.5, as
in Q1246-057, results in column density \nh$\approx~6\times10^{22}$ \cm~
and a covering fraction of $\approx$0.8.

The optical spectrum of SBS 1542+541 shows absorption troughs of
O\,VII$\lambda1032$ and Ly$\alpha$ (Fig. 2 in \citet{tel98}). From
these, we estimate the covering fraction to be 0.80, consistent with
spectral fitting results with \ax =1.5.

An ionized absorber model fit to the EPIC PN and MOS data with $N_{\rm
H,gal}$ fixed to the galactic value (1.27$\times~10^{20}$\cm) and
\ax=1.0 resulted in \nh $\approx~5.1\pm1.4\times10^{22}$ \cm~ and the
ionization state $\xi = 375 \pm 145$ (Table\ref{sbs1542_spec}), which
converts to an ionization parameter U=21.0\pl8.7.
Fig. \ref{sbs1542_contour_gam2_0} displays the contour plot between
the column density of the ionized absorber and the ionization state
$\xi$; neither parameters are very well constrained.

We noticed in the unscreened EPIC PN data of SBS 1542+541 residuals at
around 1.9 keV in the observed frame. This corresponds to the
redshifted energy of the Fe K$\alpha$ line. We used simple power-law,
partial covering model and ionized absorber models by fixing the
parameters to the results of the \ax=1.0 fits to the screened data of
SBS 1542+541. The Fe K$\alpha$ line is barely detected with 1$\sigma$
significance and we measured the equivalent widths in the range
between 70-90 eV in the observed frame corresponding to 210-270 eV in
the rest-frame.  This agrees with what is expected from Figure 3 in
\citet{nan97} for an object of log $L_{\rm 2-10 keV}$=45.8 [ergs
s$^{-1}$] (please note that here a Hubble constant of 50 km s$^{-1}$
Mpc$^{-1}$ and q$_0$=0.0 is used for comparison purposes). Because of
the low detection significance of the Fe line, however, we do not
attach much importance to this result.

We derived an X-ray loudness \aox=1.52 for SBS 1542+541 based on the
rest-frame UV spectrum given in \citet{tel98} for the flux density
$F_{\rm 2500\AA}$ at 2500\AA~ and the rest-frame 2keV flux density
calculated from an absorbed power law fit to the EPIC PN data. This
value agrees with what is expected for a radio-quiet source with an
optical luminosity density of log $l_{2500 \AA}$ = 31.6 [ergs s$^{-1}$
Hz$^{-1}$] (\citet{yua98}.

\section{\label{discuss} Discussion \& Conclusion}

The XMM-Newton observations of BAL QSOs Q1246--057 and SBS 1542+541
show that the X-ray spectra can be represented by a power law and
intrinsic partially covering absorption. The column density of the
absorber is of the order of 4$\times$10$^{22}$\cm~ with a covering
fraction of the order of 0.8. The observed column density is comparable to that
derived by \citet{tel98} to model the UV spectrum of SBS 1542+541.
A partially covering absorber and an
ionized absorber, both fit the data well. However, significant
polarization observed in Q1246--057 and non-black saturation of UV
absorption lines observed in both sources, suggests that a part of
continuum radiation is scattered back in our line of sight, favoring
the model with a partially covering absorber. It is possible that the
absorber is both warm and partially covering. In Q1246--057 the BAL
CIV and SiIV
troughs are reported to be variable (\citet{smi88}). Comparison of observations
from 1977, 1979, and 1984 found variability of the order of up to 10\%.
 The observed
X-ray variability may then be a result of change in absorber property.

The power-law slope could not be well constrained, making comparison with
spectral shape of NLS1s impossible. The absorption corrected,
intrinsic loudness of both the objects is within the normal range
observed for non-BAL QSOs. \chandra and XMM-\newton observations now offer
spectra for a handful of BAL QSOs (e.g. \citet{gal02a,gre01,
has02,cla02}). In 
Tab. \ref{balqso_xray_obs} we list 
the column densities, covering fractions and spectral slopes of published
spectra of BAL QSOs.
In many of these, a partially covering absorber is the
best fit spectral model. The detection of these sources, however, is
not only because of the part of the unabsorbed continuum in our line
of sight. The BAL QSOs detected in X-rays, and for which a reasonable
spectrum could be obtained, tend to have lower absorbing columns. As a
class, low ionization BAL QSOs tend to be more X-ray weak compared to
the high ionization BAL QSOs (\citet{gre01}), but not all hi-BALs are
the brighter members of the population either. The hi-BAL LBQS
2212-1759 is undetected in a 172 ks observation with XMM, yielding
observed \aox$\geq 2.55$ (\cite{cla02}). Thus, even the hi-BALs must
have a range of absorbing column density, plausibly related to the
strength of the CIV BAL (\cite{bra00}). 

The issue of whether BAL QSOs represent highly accreting/ younger
population of quasars remains unsettled. For the spectra presented
here, the power-law slope could not be constrained, making comparison
with spectral shape of NLS1s impossible. However, it is necessary to note that
relatively steep X-ray spectra are required for our objects (\ax$\sim$1.5) if
the covering fraction is to be the same as observed in the UV. 
[Note, however, that accretion
rate close to Eddington on to a high mass black hole may not result in
a steep X-ray power-law slope as in lower luminosity NLS1s with lower
mass black holes (see the accretion disk/ corona models of
\cite{witt97} and also \cite{joanna00})]. Comparison of rest frame
optical spectra of BAL and non-BAL quasars would be useful in deciding whether
the two classes are intrinsically similar.

\acknowledgments

We would like to thank Drs Gary Schmidt and Dean Hines for discussions
on the UV spectrum of Q1246--057, and Dr. Norbert Schartel for providing the
information about the XMM observation on LBQS 2212--1759.
This research  has made use of the NASA/IPAC Extra-galactic Database (NED)
which is operated by the Jet Propulsion Laboratory, Caltech, under
contract with the National Aeronautics and Space Administration.  The
ROSAT project is supported by the Bundesministerium f\"ur Bildung und
Forschung (BMBF/DLR) and the Max-Planck-Society. This work is
supported in part by NASA grant NAG5-9936.

\clearpage


\begin{figure*}
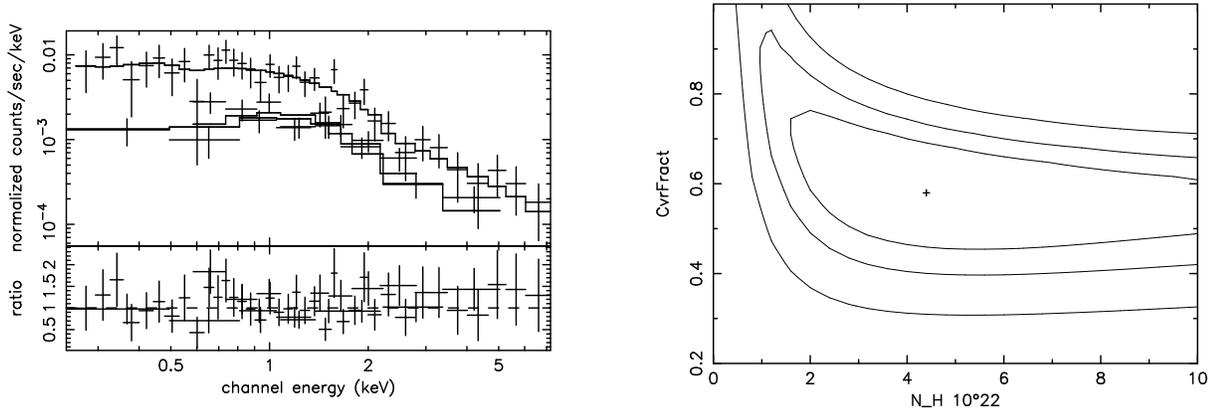

\chartlineb{DGrupe.fig1a}{DGrupe.fig1b}
\caption{\label{q1246_plot} Power-law fit with neutral galactic absorption
and intrinsic partial covering absorption to the EPIC PN, MOS-1 and
MOS-2 data of Q 1246--057 with galactic absorption fixed to the
galactic value and X-ray slope \ax~
fixed to 1.0 (see Table \ref{q1246_spec}).  The left panel shows the
fits to the spectra and the right panel displays the confidence levels
of the Column density of the partial coverer vs. the covering factor
(see Table \ref{q1246_spec}).  }
\end{figure*}

\begin{figure*}
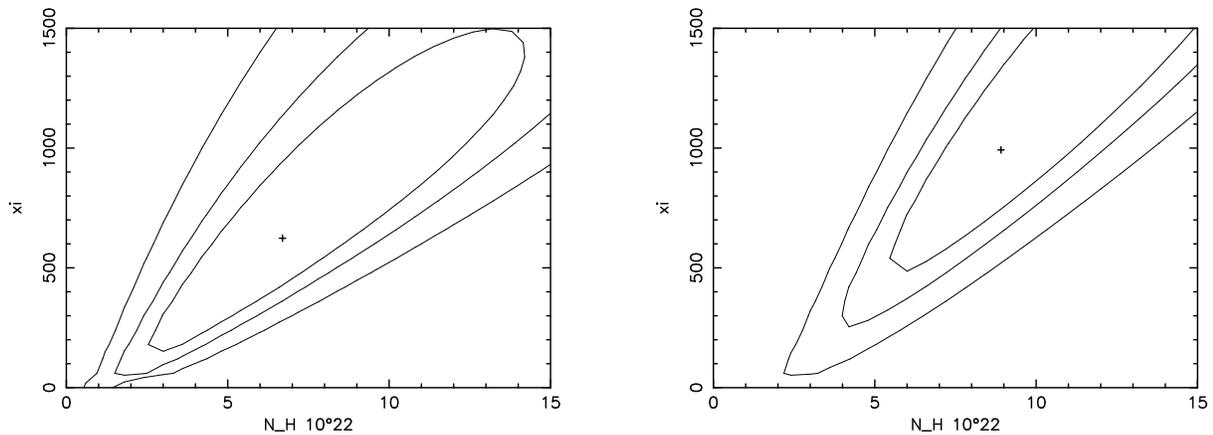

\chartlinec{DGrupe.fig2a}{DGrupe.fig2b}
\caption{\label{q1246_plot_cont} Contour plots of the absorption column and
ionization state $\xi$ using a
power-law fit with intrinsic
ionized absorber to the EPIC PN + MOS-1 and MOS-2 data of Q1246--057. The left
panel shows the contours with a fixed \ax=1.0 and the right with \ax=1.28 (see
Table \ref{q1246_spec}).
}
\end{figure*}


\begin{figure*}
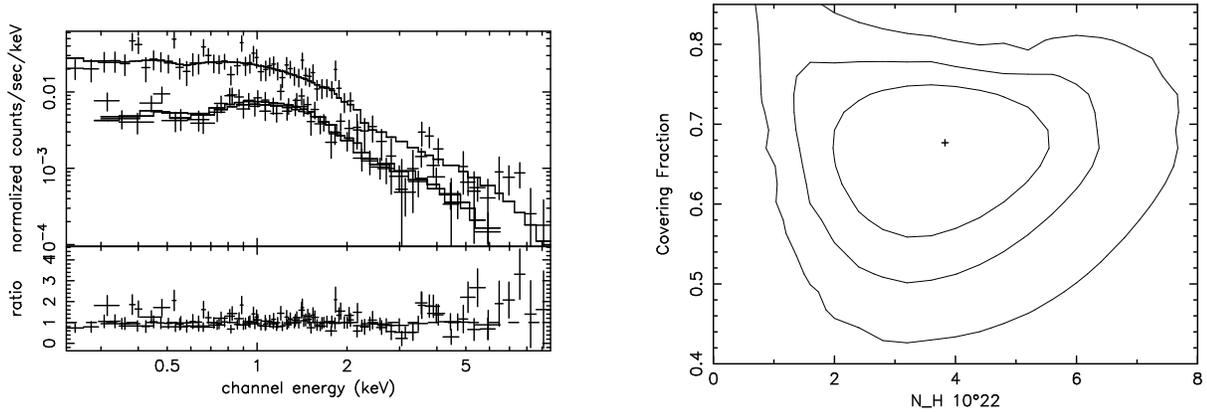

\chartlineb{DGrupe.fig3a}{DGrupe.fig3b}
\caption{\label{sbs1542_plot} Power-law fit with neutral galactic absorption
and intrinsic partial covering absorption to the EPIC PN, MOS-1, and MOS2
PSPC data of SBS 1542+541. The left panel shows the fits to the spectra and the
right panel displays the confidence levels of the Column density of the partial
coverer vs. the Photon index (see Table \ref{sbs1542_spec}).}
\end{figure*}

\begin{figure*}
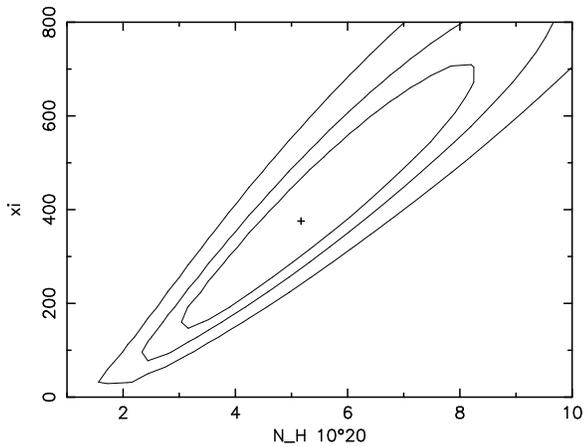

\clipfig{DGrupe.fig4}{85}{0}{0}{270}{190}
\caption{\label{sbs1542_contour_gam2_0} Contour plot of SBS 1542+541
of the column density
$N_{\rm H,intr}$ and ionization state $\xi$ of the ionized intrinsic absorber
with galactic absorption column fixed to the galactic value and \ax~=1.0.
}
\end{figure*}

\clearpage

\clearpage

\begin{deluxetable}{lcrccr}
\tabletypesize{\scriptsize}
\tablecaption{Summary of the XMM observations of Q1246--057 and SBS1542+541
 \label{xmmobs}}
\tablewidth{0pt}
\tablehead{
\colhead{Source} & \colhead{Obs ID}   & \colhead{Instr.}   &
\colhead{$\rm Obs_{start}$} &
 \colhead{$\rm Obs_{end}$} & \colhead{$T_{obs}$} 
}
\startdata
Q1246--057 & 0060370201 & PN & 2001-07-11, 15:59:38 & 2001-07-12, 02:40:56 &  
38008 \\
& & MOS & 2001-07-11, 15:21:40 & 2001-07-12, 02:39:54 & 40514 \\ \\
SBS 1542+541 & 0060370101 & PN & 2002-02-03, 04:15:32 & 0220-02-03, 06:45:48 &
8546\tablenotemark{1} \\
& & MOS & 2002-02-03, 03:41:54 & 2002-02-03, 06:47:05 & 10822\tablenotemark{1} 
\\ \\
& 0060370901 & PN & 2002-02-06, 20:57:59 & 2002-02-07, 06:31:35 & 
33946\tablenotemark{1} \\
& & MOS & 2002-02-06, 20:24:21 & 2002-02-06, 06:31:03 & 36222\tablenotemark{1} 
\\
\enddata

\tablenotetext{1}{Please note that parts of the observations could not be used
for spectral analysis due to high background radiation. The actual times used 
are
given in Sect. \ref{xmm_observe}.
}

\end{deluxetable}

\begin{deluxetable}{lcccccr}
\tabletypesize{\scriptsize}
\tablecaption{Spectral Fit parameters of Q1246--057 for a power-law
model with galactic absorption ($N_{\rm H, gal}=2.15~10^{20}$ \cm) and 
red-shifted partial coverer and ionized absorption for the intrinsic 
absorption.
The ionization state is given by $\xi$ (\citet{don92}).
 \label{q1246_spec}}
\tablewidth{0pt}
\tablehead{
&  \colhead{$N_{\rm H, gal}$} &  \colhead{$N_{\rm H, intr}$} & 
\colhead{Covering} & \\
\colhead{\rb{Detector}} & \colhead{10$^{20}$\cm}  & \colhead{10$^{22}$\cm}  
 & \colhead{Fraction}   &
\colhead{\rb{$\xi$}} & \colhead{\ax} & \colhead{\rb{$\chi^2$ (DOF)}} 
}
\startdata
PN\tablenotemark{1}  & 2.15 (fixed) & --- & --- & --- & 0.794\pl0.314 & 6.9 
(12) \\
  &  92.0\pl157 & --- & --- & --- & 1.28\pl0.94 & 7.5 (11) \\ \\
PN\tablenotemark{2} & 4.78\pl3.50 & --- & --- & --- & 0.710\pl0.213 & 22.1 
(34) \\
   & 2.15 (fixed) & 2.11\pl5.07 & 0.42\pl0.30 & --- & 0.780\pl0.273 & 21.5 
(33)  \\
   & 2.15 (fixed) & 2.47\pl3.62 & 0.43\pl0.27 & --- & 0.800 (fixed) & 21.6 
(33)  \\
   & 2.15 (fixed) & 2.55\pl1.09 & 0.86\pl0.09 & --- & 1.280 (fixed) & 30.8 (35)
    \\
   & 2.15 (fixed) & 5.28\pl3.92 & --- & 495\pl470 & 1.000 (fixed) & 21.8 (34)
    \\ \\
MOS-1\tablenotemark{3} & 6.41\pl7.85 & --- & --- & --- & 0.537\pl0.339 & 4.1 
(6) \\
    & 2.15 (fixed) & 15.05\pl9.13 & 0.80\pl0.21 & --- &  1.216\pl0.564 & 2.2 
(5) \\
    & 2.15 (fixed) & 10.60\pl11.39 & 0.60\pl0.15 & --- & 0.800 (fixed) & 3.2 
(6) \\ \\
MOS-2\tablenotemark{4} &  3.70\pl10.95 & --- & --- & --- & 0.642\pl0.652 & 6.3 
(5)
\\
    & 2.15 (fixed) & 0.49\pl17.59 & 0.95\pl800 & --- & 0.800\pl0.874 & 6.0 (4) 
\\ \\
PSPC\tablenotemark{5} (0.2-1.45 keV)
    & 2.15 (fixed) & 4.56\pl11.78 & 0.95\pl14.25 & --- & 1.520\pl5.450 & 3.5 
(7) \\
    & 2.15 (fixed) & 2.84\pl4.00 & 0.95\pl8.08 & --- & 1.000 (fixed) & 3.2 (8) 
\\
    & 2.15 (fixed) & 3.90\pl3.86 & 0.95\pl3.32 & --- & 1.500 (fixed) & 3.6 (8) 
\\ \\
MOS-1\tablenotemark{3} + MOS-2\tablenotemark{4}
 & 3.65\pl5.82 & --- & --- & --- & 0.510\pl0.300 & 11.4 (13) \\
    & 2.15 (fixed) & 0.47\pl13.50 & 0.48\pl10.8 & --- & 0.511\pl0.320 & 11.1 
(13) \\
    & 2.15 (fixed) & 4.170\pl540 & 0.48\pl0.19 & --- & 0.800 (fixed) & 11.7 
(13) \\
    & 2.15 (fixed) & 11.12\pl4.90 & 0.74\pl0.06 & --- & 1.280 (fixed) & 14.9 
(14) \\ \\
PN\tablenotemark{2} + MOS-1\tablenotemark{3} + MOS-2\tablenotemark{4} &  
5.52\pl3.48 & --- & --- & --- & 0.723\pl0.176 & 31.8 (46)
\\
    & 2.15 (fixed) & 2.49\pl4.70 & 0.38\pl0.29 & --- & 0.730\pl0.222 & 33.2 
(46) \\
    & 2.15 (fixed) & 2.76\pl3.56 & 0.44\pl0.17 & --- & 0.800 (fixed) & 33.9 
(47) \\
    & 2.15 (fixed) &  4.49\pl2.69 & 0.58\pl0.09 & --- & 1.000 (fixed) & 35.3 
(47) \\
    & 2.15 (fixed) & 6.08\pl2.09 & 0.72\pl0.06 & --- & 1.280 (fixed) & 39.9 
(48) \\
    & 2.15 (fixed) & 7.43\pl2.07 & 0.79\pl0.04 & --- & 1.500 (fixed) & 44.2 
(47)
     \\ 
    & 2.15 (fixed) & 3.60\pl3.39 & --- & 445\pl560 & 0.800 (fixed) & 33.0 (47)
    \\
    & 2.15 (fixed) & 6.69\pl10.51 & --- & 630\pl1390 & 1.000 (fixed) & 34.5 
(47)
    \\
    & 2.15 (fixed) & 9.11\pl2.95 & --- & 990\pl435 & 1.280 (fixed) & 38.9 (47)
    \\ \\
PN\tablenotemark{2} + MOS-1\tablenotemark{3} + MOS-2\tablenotemark{4}  + 
PSPC\tablenotemark{5}  & 4.98\pl3.26 & --- & --- & --- &
0.685\pl0.171 & 34.7 (51) \\
 & 2.15 (fixed) &  0.33\pl1.00 
 & 0.95\pl56.2 & --- & 0.644\pl0.155   & 39.3 (54) \\
 & 2.15 (fixed) & 0.64\pl0.62 & 0.95\pl1.75 & --- & 0.800 (fixed) & 40.4 (55) 
\\
 & 2.15 (fixed) & 5.65\pl1.98 & 0.73\pl0.05 & --- & 1.280 (fixed) & 41.5 (54) 
\\
 & 2.15 (fixed) & 3.74\pl1.72 & --- & 430\pl200 & 0.800 (fixed) & 34.9 (53) \\
 & 2.15 (fixed) & 6.58\pl2.35 & --- & 590\pl240 & 1.000 (fixed) & 36.3 (53) \\
 & 2.15 (fixed) & 9.10\pl2.67 & --- & 985\pl370 & 1.280 (fixed) & 41.3 (53) \\ 
\enddata

\tablenotetext{1}{EPIC PN, observed energy range used: 1.5-7.5 keV}
\tablenotetext{2}{EPIC PN, observed energy range used: 0.2-7.5 keV}
\tablenotetext{3}{EPIC MOS-1, observed energy range used: 0.2-7.5 keV}
\tablenotetext{4}{EPIC MOS-2, observed energy range used: 0.2-4.0 keV}
\tablenotetext{5}{ROSAT Position Sensitive Proportional Counter, 
observed energy range used: 0.2-1.45 keV}
\end{deluxetable}

\begin{deluxetable}{lrr}
\tabletypesize{\scriptsize}
\tablecaption{Redshifts, distances (in Mpc), 
unabsorbed rest-frame X-ray Luminosities (in units of ergs s$^{-1}$) 
the rest-frame luminosity density $l_{2500 \AA}$  at 2500 \AA~ (in units of 
ergs
s$^{-1}$ Hz$^{-1}$,
and X-ray loudness \aox\tablenotemark{1}
 \label{xraylum}
 }
\tablewidth{0pt}
\tablehead{
\colhead{Property} & \colhead{Q1246--057}   & \colhead{SBS 1542+541}  
}
\startdata
z & 2.236 & 2.361 \\
D & 11500 & 12222 \\
log $L_{\rm X}$ (0.2-2.0 keV) & 44.70 & 45.13 \\
log $L_{\rm X}$(2.0-10.0 keV) & 44.64 & 45.10 \\
log $l_{2500 \AA}$ & 31.08 & 31.62 \\
\aox & 1.50\tablenotemark{2} & 1.48\tablenotemark{2}
\enddata

\tablenotetext{1}{The X-ray loudness is defined by: \ax~=~--0.384 log($l_{\rm
2keV}/l_{\rm 2500 \AA}$).  }
\tablenotetext{2}{Assuming partial covering model}

\end{deluxetable}

\begin{deluxetable}{lcccccr}
\tabletypesize{\scriptsize}
\tablecaption{Spectral Fit parameters of SBS1542+541 for a power-law
model with galactic absorption ($N_{\rm H, gal}=1.27~10^{20}$ \cm) and
red-shifted partial coverer or ionized absorption for the intrinsic 
absorption.
 \label{sbs1542_spec}}
\tablewidth{0pt}
\tablehead{
&  \colhead{$N_{\rm H, gal}$} &  \colhead{$N_{\rm H, intr}$}
& \colhead{Covering} & \\
\colhead{\rb{Detector}} & \colhead{10$^{20}$\cm} & \colhead{10$^{22}$\cm}
  & \colhead{Fraction}   & \colhead{\rb{$\xi$}} &
\colhead{\ax} & \colhead{\rb{$\chi^2$ (DOF)}} 
}
\startdata
PN\tablenotemark{1} & 1.27 (fixed) & --- & --- & --- & 1.126\pl0.220 & 
30.7 (26) \\
PN\tablenotemark{1} & 0.00\pl1.70 & --- & --- & --- & 1.044\pl0.574 & 31.0
(25) \\  \\
PN\tablenotemark{2} & 8.43\pl1.98 & --- & --- & --- & 0.965\pl0.119 & 
68.0 (67)
\\
 & 1.27 (fixed) & --- & --- & --- & 0.581\pl0.048 & 91.2 (68) \\
 & 1.27 (fixed) & 3.58\pl1.26 & 0.73\pl0.08 & --- & 1.129\pl0.161 & 64.3 
(66) \\
 & 1.27 (fixed)  & 5.52\pl0.83 & 0.84\pl0.02 & --- &  1.500 (fixed) & 
72.3 (67)
  \\
 & 1.27 (fixed) & 3.98\pl1.45 & --- & 250\pl140 & 1.000 (fixed) & 66.5 
(67) \\
 & 1.27 (fixed) & 7.98\pl1.41 & --- & 1000\pl280 & 1.500 (fixed) & 76.5 
(67) \\
 \\
 MOS-1\tablenotemark{3} & 10.91\pl4.10 & --- & --- & --- & 0.962\pl0.188 
& 27.4
 (26) \\
 & 1.27 (fixed) & 3.48\pl2.78 & 0.68\pl0.14 & --- & 1.000\pl0.223 & 26.9 
(25) \\
   & 1.27 (fixed) & 6.34\pl1.52 & 0.83\pl0.04 & --- & 1.500 (fixed) & 
33.3 (26) \\
   & 1.27 (fixed) & 7.08\pl3.90 & --- & 570\pl440 & 1.000 (fixed) & 27.1 
(26 \\
   \\
MOS-2\tablenotemark{4} & 6.50\pl3.76 & --- & --- & --- & 0.813\pl0.181 & 
21.3 (23)
\\
   & 1.27 (fixed) & 6.35\pl4.70 & 0.55\pl0.20 & --- & 0.984\pl0.266 & 
20.0 (22) \\
   & 1.27 (fixed) & 10.67\pl3.09 & 0.80\pl0.04 & --- & 1.500 (fixed) & 
26.8 (23) \\
   & 1.27 (fixed) & 9.65\pl5.25 & --- & 1000\pl620 & 1.000 (fixed) & 
20.4 (23)
   \\ \\
MOS-1\tablenotemark{3} + MOS-2\tablenotemark{4} & 8.89\pl2.83 & --- & 
--- & ---
& 0.894\pl0.132 & 49.2 (51) \\
& 1.27 (fixed) & 4.18\pl2.50 & 0.60\pl0.12 & --- & 0.962\pl0.167 & 47.9 
(50) \\
& 1.27 (fixed) & 10.37\pl3.20 & --- & 1000\pl390 & 1.124 (fixed) & 49.2 
(51) \\
\\
PN\tablenotemark{2} + MOS-1\tablenotemark{3} + MOS-2\tablenotemark{4}
 & 8.43\pl1.64 & --- & --- & --- & 0.923\pl0.088 & 118.3 (120) \\
 & 1.27 (fixed) & 3.80\pl1.13 & 0.68\pl0.06 & --- & 1.053\pl0.111 & 
113.7 (119) \\
 & 1.27 (fixed) & 6.33\pl0.72 & 0.83\pl0.02 & --- & 1.500 (fixed) & 
131.0  (120) \\
 & 1.27 (fixed) & 5.17\pl1.40 & --- & 377\pl150 & 1.000 (fixed) & 116.5 
(120) \\
\enddata

\tablenotetext{1}{EPIC PN, observed energy range used: 1.5-10.0 keV}
\tablenotetext{2}{EPIC PN, observed energy range used: 0.2-12.0 keV}
\tablenotetext{3}{EPIC MOS-1, observed energy range used: 0.2-9.0 keV}
\tablenotetext{4}{EPIC MOS-2, observed energy range used: 0.2-7.0 keV}

\end{deluxetable}

\clearpage

\begin{deluxetable}{lccccl}
\tabletypesize{\scriptsize}
\tablecaption{Summary of published Chandra and XMM observation of BAL QSOs
 \label{balqso_xray_obs}}
\tablewidth{0pt}
\tablehead{
\colhead{Source} & \colhead{Mission}   & \colhead{Intrinsic
$N_H$\tablenotemark{1} }  &
\colhead{Covering Fraction} & \colhead{\ax} &
 \colhead{reference} \\ 
}
\startdata
Q1246--057 & XMM, EPIC PN & 2.47\pl3.62  &  0.43\pl0.27 & 0.800 & this paper \\
SBS 1542+541 & XMM, EPIC PN & 3.58\pl1.26 & 0.73\pl0.08 & 1.129 & this paper \\
\\
Q 0000--263 & XMM, PN+MOS & 0.0 & --- & 1.1 & \citep{fer03} \\
APM 08279+5255 & Chandra, ACIS-S & 6.0\pl3.5 & --- & 0.86 & \citep{cha02} \\
APM 08279+5255 & XMM, PN + MOS & 6.9\pl0.3 & --- & 1.04 & \citep{has02} \\
RX J0911.4+0551 & Chandra, ACIS-S & 19\pl28 & 0.71\pl0.20 & 0.87 & \citep{cha01}
\\
PG 1115+080 & Chandra, ACIS-S & 3.8\pl2.5 & 0.64\pl0.12 & 0.99 & \citep{cha02}
\\
Mkn 231 & Chandra, ACIS-S & 2.1$^{+1.3}_{-0.9}$ & --- & 1.1 & \citep{gal02b} \\
H 1413+117 & Chandra, ACIS-S & 20\pl18 & --- & 0.39 & \citep{cha02} \\
PG 1411+442 & Chandra, ACIS-S & 19\pl7.5 & 0.97\pl0.02 & 1.20 & \citep{gal02a}
\\
PG 1535+547 & Chandra, ACIS-S & 12\pl8 & 0.91\pl0.07 & 1.02 & \citep{gal02c} \\
PG 2112+059 & Chandra, ACIS-S & 1.1\pl0.5 & --- & 0.97 & \citep{gal02c} \\
LBQS 2212--1759 & XMM, EPIC PN & $>$3000 & --- & 0.9 & \citep{cla02} \\
PHL 5200 & XMM, EPIC PN & 10 & --- & 1.0 & \citep{bri02} \\
\\
\enddata

\tablenotetext{1}{In units of 10$^{22}$ \cm}

\end{deluxetable}

\end{document}